# Implementation of distillation protocols using a recirculating "bricks" mesh network

Jacek Gosciniak

*Institute of Microelectronics and Optoelectronics, Warsaw University of Technology,*

*Koszykowa 75 st., Warsaw 00-662, Poland*

Email: jacek.gosciniak@pw.edu.pl

**Abstract**

General-purpose programmable photonic processors provide a flexible foundation for integrating various functionalities within a single chip. A two-dimensional "bricks" waveguide mesh of Mach–Zehnder interferometers has been demonstrated to possess considerable potential in the domain of photonic neural networks and quantum signal processing.

In this article, we propose an expansion of the available applications of recirculating "bricks" mesh architecture to distillation protocols necessary for quantum signal processing. These protocols are essential for the heralding of the output of single photons, which is characterized by a reduced distinguishability error rate. The demonstration will be made of a single programmable optical system's ability to realize various distillation protocols with reduced computational resource costs. The present study will concentrate on cascaded quantum interferometers and Fourier transform-based schemes. It will demonstrate that the "bricks" mesh can implement such schemes, which are unattainable using feed-forward networks, without the need for complex out-of-plane integration. The propagation of the signal in any direction, along with the utilization of all ports as both input and output, facilitates the execution of such transformations with minimal optical depth of the circuit and in time scales shorter than the decoherence time.

## Introduction

The rapid progress of quantum computing in recent years has been predominantly driven by the prospect of computational advantage, defined as the ability to solve problems more rapidly than classical computers [**1, 2**]. The development of this promising technology hinges upon the generation and manipulation of multiple indistinguishable photons, which are essential for the implementation of effective photonic gates [**3-5**]. Integrated photonics has emerged as a promising hardware candidate for these tasks, and universal fault-tolerant quantum computing has been identified as a critical component for scalable photonic quantum information processing technologies [**6, 7**]. At the core of this technology is a network comprising multiple beam splitters and phase shifters, which form multiport interferometers [**6, 8**]. These multiport interferometers provide precise control over quantum states of light, and their implementation as photonic integrated circuits (PICs) enhances resilience to environmental noise, which is critical for high-fidelity, coherent quantum state manipulation. It is imperative that such circuits be characterized by universal reconfigurability, which is the ability to program a device to realize any unitary transformation between input and output modes. The core of the programmable PIC's processor is constituted by a photonic waveguide mesh, a two-dimensional lattice that provides regular and periodic geometry, formed by replicating unit cells [**9-13**]. This objective is typically realized through the implementation of feed-forward architectures such as the triangular mesh architecture proposed by Reck [**14**] or the rectangular mesh architecture proposed by Clements [**15**]. Nevertheless, the flow of light within these networks is constrained to a single direction [**16-18**], thereby impeding the versatility of the systems. To expand the range of signal processing capabilities, there is a necessity to transition to networks that offer a wider range of possibilities. A recirculating mesh offers notable advantages, chiefly in terms of scalability, high reconfigurability, and reduced optical losses when compared to conventional feed-forward mesh architectures [**10, 12, 19**]. Rather than constructing a massive, one-way interferometers, a recirculating mesh permits photons to propagate through a smaller, programmable, and tunable component multiple times to simulate a larger, complex unitary transformation.

**Recirculating “bricks” mesh architecture**

The recirculating "bricks" mesh architecture consists of two to four MZIs in a unit cell [**19-21**], compared to six for a hexagonal mesh [**9, 10, 12, 13**]. This results in a significant reduction in the optical path length and, consequently, the propagation losses [**19**]. Concurrently, it produces a more efficient 3-point interconnection scheme in comparison to a conventional square mesh architecture, wherein each unit cell is connected through four points. The capacity for signal flow in any direction facilitates the implementation of both control loops and various types of filters, including both infinite impulse response (IIR) filters, which are based on ring resonators (RRs), and finite impulse response (FIR) filters, which are based on asymmetric MZIs. This outcome signifies a substantial progression beyond the functional limitations of feed-forward networks, thereby paving the way for the integration of more intricate systems [**19-21**].

The reconfiguration performance of the "bricks" mesh, defined as the number of filters with different frequency separation values for the RR-based filter, amounts to 11 for only 25 MZIs. In comparison, for a hexagonal and triangular mesh with the same number of MZIs, the calculations yielded 9 and 6, respectively, while for a regular square mesh, the calculation resulted in 6 [**19**]. Concurrently, the reconfiguration performance for the MZI-based filter configured within the "bricks" mesh architecture attains a value of 12, which is equivalent to the performance of the hexagonal mesh. In comparison, the calculation for a triangular mesh yielded a value of 8, while for a regular square mesh, the result was 6 [**19**]. As previously mentioned, the "bricks" mesh architecture exhibits a twofold enhancement in performance in comparison to conventional square mesh architecture [**9**]. This is yet another undeniable advantage of an architecture built on a "bricks" mesh.

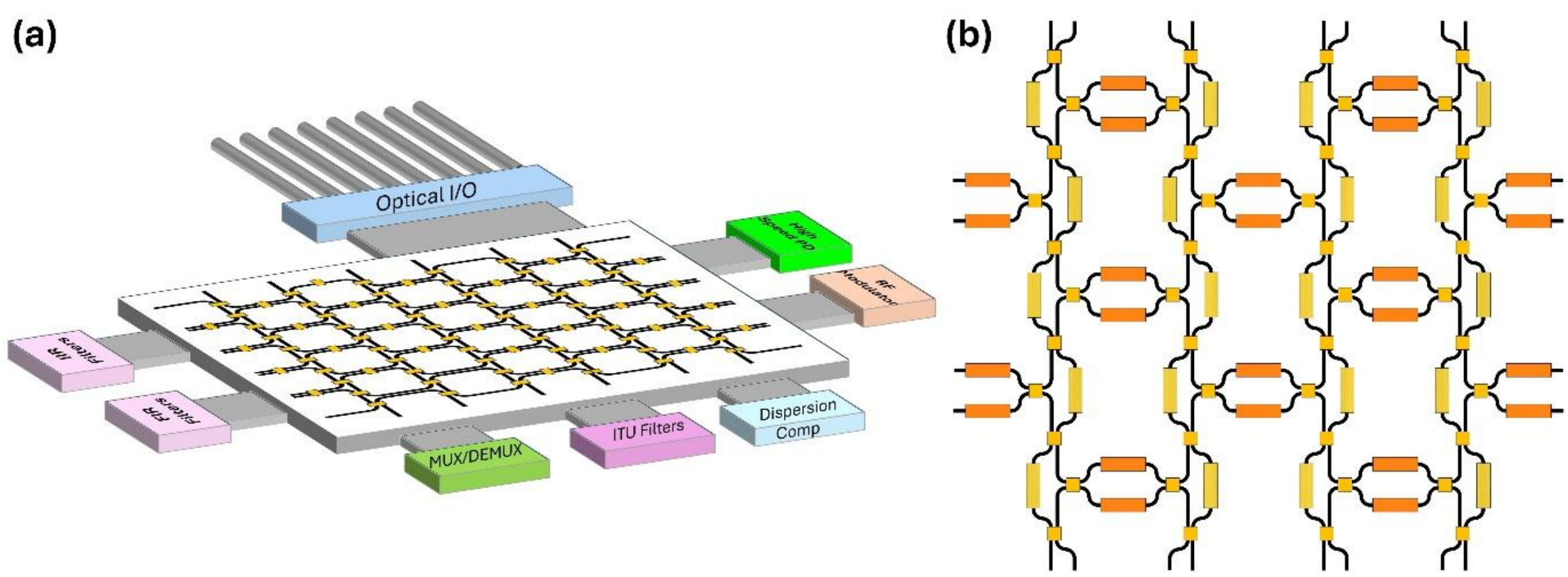

**Figure 1**. (a) The photonic chip with a (b) programmable mesh core connected to control electronics, optical fibers, filters, high-speed modulators and detectors and many other components

As demonstrated in previous studies [**20**], the optical depth of a network based on the "bricks" mesh architecture is significantly reduced in comparison to feed-forward networks, both based on rectangular and triangular meshes [**9**]. Furthermore, the number of active MZIs required to perform any linear transformation between input and output modes is reduced by more than a factor of 13. Additionally, all ports capable of functioning as both input and output ports are situated on all four sides of the mesh structure [**19-21**]. This facilitates the implementation of systems that are not achievable with feed-forward networks. Concurrently, the recirculating "bricks" mesh architecture bears a striking resemblance to feed-forward networks, a characteristic that facilitates the execution of all conceivable tasks associated with this category of network.

A proposed recirculating "bricks" mesh architecture exhibits another advantage in its compatibility with a recently proposed monitoring system that can monitor signal flow in each part of the circuit under operation [**22**]. The system is able to execute a control strategy that automatically adjusts the optical power and phase of photonic components at each point of the photonic system with extreme accuracy and minimal insertion loss. The technique is predicated on a feedback control loop that concurrently adjusts the matrix coefficient of the device transfer function and compensates for process tolerances and thermal drift in real time. The control strategy utilizes a Wheatstone bridge configuration with a feedback loop that operates without the requirement of prior knowledge concerning the device transfer function [**22**].

The system can be constructed using transparent conductive oxides (TCOs), which are defined by the epsilon-near-zero (ENZ) point, and whose properties can be actively modified by an external voltage [**23, 24**].

**Input-output configurations**

The implementation of a recirculating "bricks" mesh architecture allows for a wide variety of input-output configurations, paving the way for various types of transformations realized between the system's input and output that cannot be achieved with a feed-forward network. Examples of various input-output system implementations are shown in **Fig. 2**.

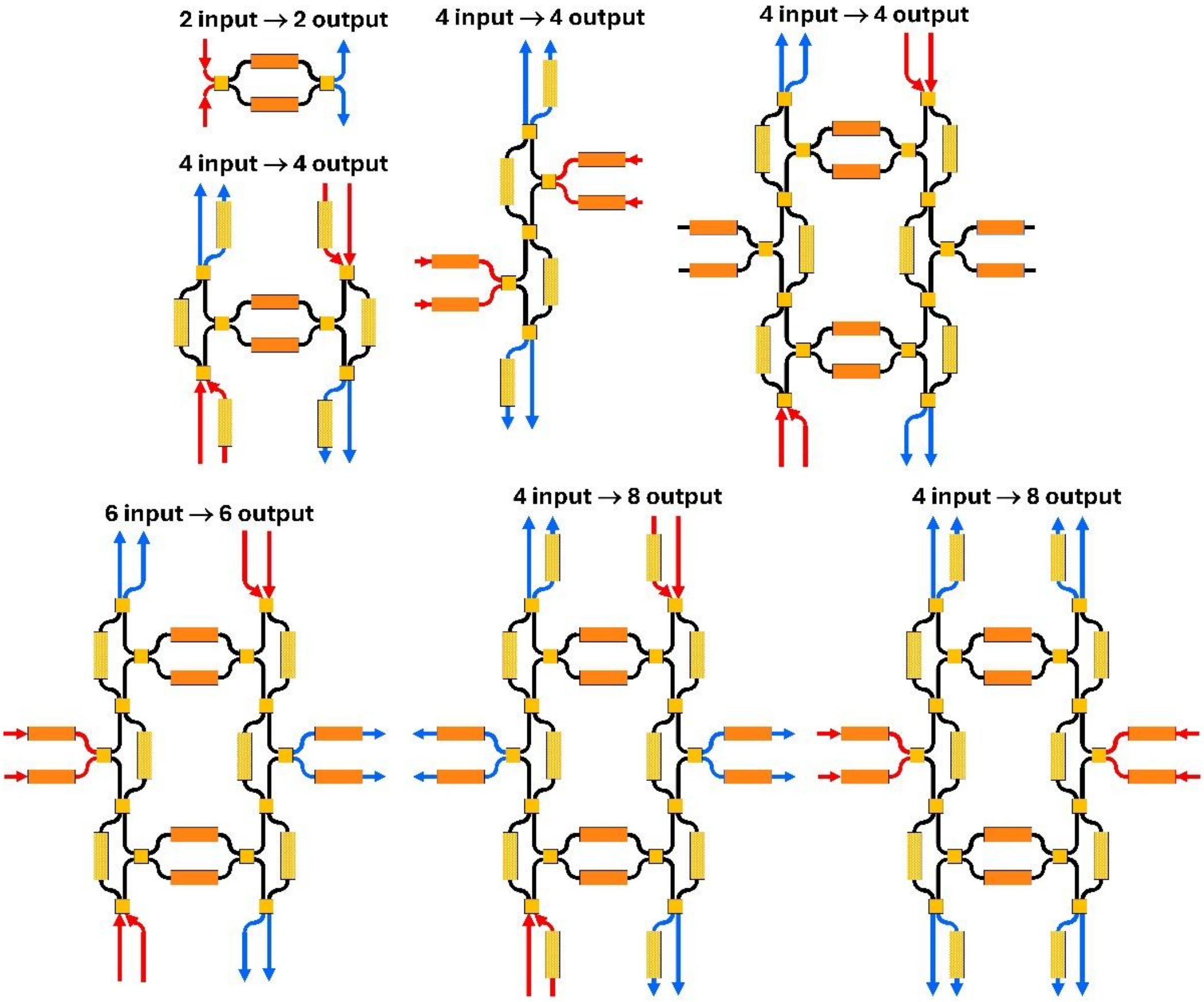

**Figure 2**. Examples of various input-output system implementations realized in the recirculating “bricks” mesh architecture.

The capacity to select an appropriate unitary transformation between the input and output of the system exerts a significant influence on the quantum interference effects present in multimode interferometers. This, in turn, has the potential to result in a substantial suppression of output configurations, which consequently leads to a notable reduction in computational resource costs [**4, 25,26**].

**Photon indistinguishability**

Photon indistinguishability stems from the fundamental properties of light, in which, on the one hand, linearity precludes explicit photon-photon interactions, and on the other hand, it enables multi-particle quantum interference, which is sufficient for universal quantum computation [**3, 27-29**]. To perform such tasks, it is necessary that individual photons be indistinguishable from one another in terms of any of their internal degrees of freedom, such as polarization, frequency, or arrival time. The basic scheme for determining the degree of photon indistinguishability is based on Hong–Ou-Mandel (HOM) interference [**30**]. In the context of a balanced beam splitter, when two photons enter distinct input ports, full indistinguishability results in perfect bunching, whereby the photons invariably emerge from the splitter in the same port. In contrast, if the photons are fully distinguishable, that is, if they possess orthogonal internal states, they do not interfere, and, as would be expected for classical particles, they exit opposite ports with probability 1/2. Photons that are partially distinguishable are located between these two limits [**3, 27-29**]. The degree of two-photon indistinguishability is quantified by the visibility $V_{12}$, which is defined as the reduction of coincidence counts relative to the classical case. In the context of photonic quantum computing, photon indistinguishability emerges as a pivotal resource, with imperfect indistinguishability resulting in logical errors during computation.

**Distillation protocol**

In view of the aforementioned arguments, a primary challenge in implementing quantum computations using photons is enhancing the degree of indistinguishability. This objective is accomplished by employing photonic distillation schemes, which are non-deterministic methods for heralding photons with a reduced indistinguishability error [**27-29**]. These schemes leverage interference and measurement to exchange multiple weakly indistinguishable photons for a single photon with reduced distinguishability error. In contradistinction to the process of optical filtering, photon distillation does not necessitate prior knowledge about the target wave function of indistinguishable photons, and filters all internal degrees of freedom concomitantly.

Apart from the distinguishability of photons, a significant challenge in quantum information processing is the realization of key transformations on time scales shorter than the decoherence time and with reduced losses, as both factors influence the distinguishability of photons. The photon losses mainly arise from the propagation losses within the circuit and are closely dependent on the system architecture. They are determined by the optical depth of the chip, thus, depending on the number of components the photons have to go through and the loss per Mach-Zehnder interferometer (MZI).

This assertion extends to the decoherence time, whose significance rises concomitantly with the increase in the optical depth of the circuit. It is imperative to note that both of these factors have a substantial impact on the distinguishability of photons, thereby underscoring the crucial necessity of minimizing the network size for the purpose of enhancing the efficacy of quantum computing.

This can be realized through an architecture-based approach using the recirculating photonic networks that minimize the duration of information processing tasks through dynamic control of multi-photon interactions. It has been shown that recirculating photonic networks may provide significant improvements in time and hardware efficiency relative to state-of-the-art architectures based on feed-forward networks [**32**].

**Cascaded HOM interferometers**

The objective of the distillation protocol is to herald the output of single photons with reduced distinguishability error rate, thereby enabling more efficient linear optical quantum computation. The process itself is realized through linear optical mechanism in which generated photons are processes with a linear optical interferometer to purify single indistinguishable photons. In the field of quantum information processing, it involves the consumption of multiple copies of noisy quantum states to obtain a single output state in which the noise is suppressed [**31**].

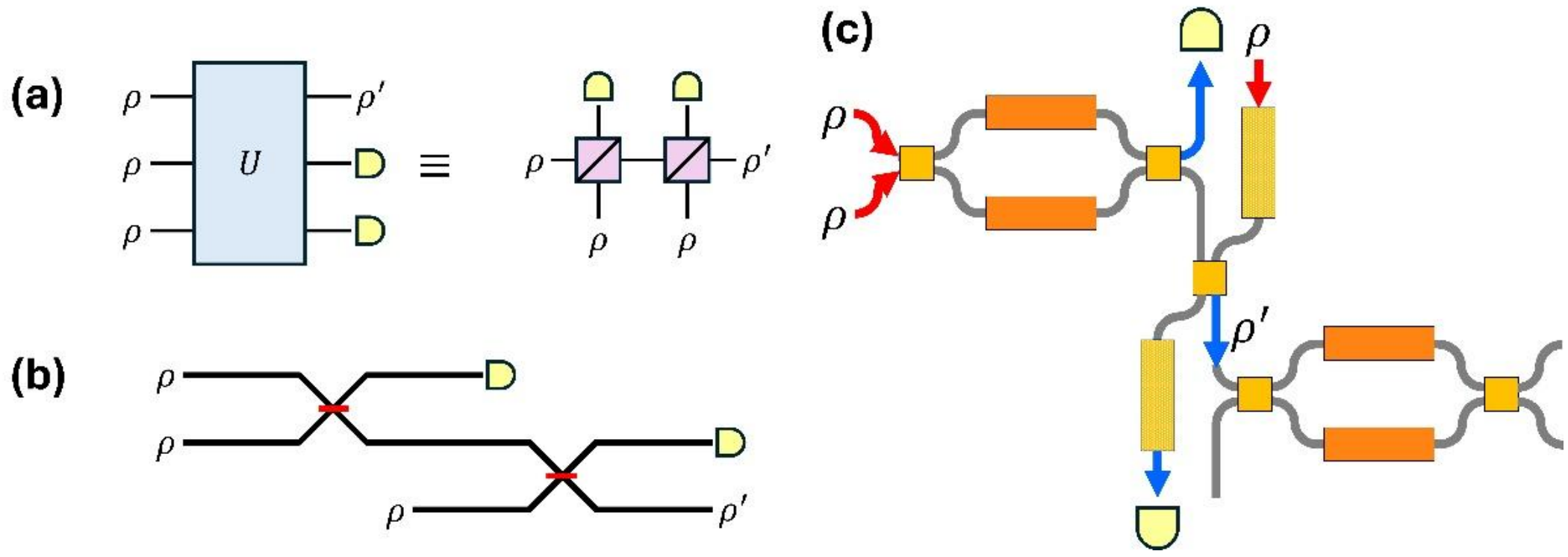

**Figure 3**. (a) Schematic of the photon distillation gate for the raw of input photonic $\rho$ and the distilled output photonics $\rho'$. (b) Linear optical circuit, based on cascaded Hong-Ou-Mandel-type quantum interferometers, for purifying indistinguishable $n = 2$ photons.

The example of simple distillation gate is shown in **Fig. 3a** where the input state $\sigma = \rho(\epsilon)^{\otimes n}$ consisting of $n = 3$ copies of noisy photon state are used to produce single photon with reduced degree of distinguishability. Input photons populate spatial modes, while the box is a circuit composed of beam splitters and phase shifters. The output photon is conditioned on the post-selection of a particular measurement outcome, i.e., the detection of $n - 1$ photons in some configuration. This scheme relies on the fundamental property of photons, which is that identical photons interfere in a fundamentally different manner than partly distinguishable ones and, may be used to reduce the distinguishability of noisy photon sources [**27**].

**Fig. 3b** shows a block diagram illustrating the beam splitters and waveguides of standard, programmable linear optical circuit to measure the indistinguishability error in the presence of other noise sources that implement the distillation protocol on two photons. The third photon is then used as a reference to measure the distinguishability via Hong-Ou-Mandel (HOM) interference [**30**]. Such protocols take advantage of the disparity in the statistics of indistinguishable and distinguishable photons interfering on a beam splitter to amplify the indistinguishable components of the quantum state [**27**, **33**]. In accordance with the Hong-Ou-Mandel (HOM) effect, indistinguishable photons bunch when interfering on a balanced beam splitter (BS), while distinguishable photons only do so half of the time. Therefore, the absence of a detected photon in a single output port of the beam splitter (signaling the presence of two bunched photons in the other mode) results in an enhancement of the amplitude of the indistinguishable component of the photons' state, as the distinguishable component is less likely to produce such a measurement event. The subsequent extraction of a purified single photon from a pair of bunched photons is achieved through a heralded process involving probabilistic splitting with an additional BS, followed by detection of a single photon in one of the output arms.

This type of operation can be classified as partial-measurement operation (that refers to the projective measurements on one or more output modes) where only a part of modes is measured while quantum information is left in the remaining mode for further processing [**28**]. In most cases, the nonlinearity required for universal photonic quantum information processing is introduced through feed-forward operations in linear optical circuits. The linear optical circuit is able to perform any arbitrary unitary transformation on a set of optical modes by decomposing it into sequence of unitary operations acting on a two-mode subspace. Each such unitary operation can be performed by a combination of beam splitters and phase shifters arranged in different configurations (**Fig. 3b**). Usually, it consists of MZI with two phase shifters and two 50:50 beam splitters. In the case of symmetric MZI (sMZI) the phase shifters are arranged in both arms of the interferometer while the asymmetric MZI (aMZI) consists of one internal and one external phase shifter. The linear optical circuit that implements one of the above basic components can be arranged in the triangular design, proposed by Reck [**14**], or the rectangular design, proposed by Clements [**15**], which differ in optical depth and loss asymmetry. Thereby, a combination of an optical transformation with a partial measurement which projects onto a single photon state ($\rho(\epsilon')$), results in photon distillation (when $\epsilon' < \epsilon$).

The circuit presented in **Fig. 3a** is able to achieve 2.2-fold reduction in the indistinguishability error for photons after distillation compared to photons before distillation gate. The indistinguishability errors can be emulated through the use of a series of beam splitters implemented via a feed-forward-type network, as illustrated in **Fig. 3b**. In the first beam splitter, the interaction between the signal photon and the noise photon occurs, thereby introducing multiphoton noise, while in the second beam splitter, the loss error is modeled through the interaction between the monitored output mode and an auxiliary mode.

The same task can be accomplished using a recirculating "bricks" network, which allows for signal propagation not only horizontally but also vertically (**Fig. 3c**). This development enables the network to perform a multitude of tasks that are challenging to achieve with feed-forward networks. A comparison of the two schemes illustrated in **Fig. 3b and 3c**, respectively, clearly demonstrates that the distillation protocol employing a recirculating "bricks" mesh architecture (**Fig. 3c**) attains a twofold reduction in network depth in comparison to a feed-forward network (**Fig. 3b**). Going into details, for $n = 2$ photons, a feed-forward network necessitates two MZI layers, whereas the "bricks" mesh requires only one layer of MZI.

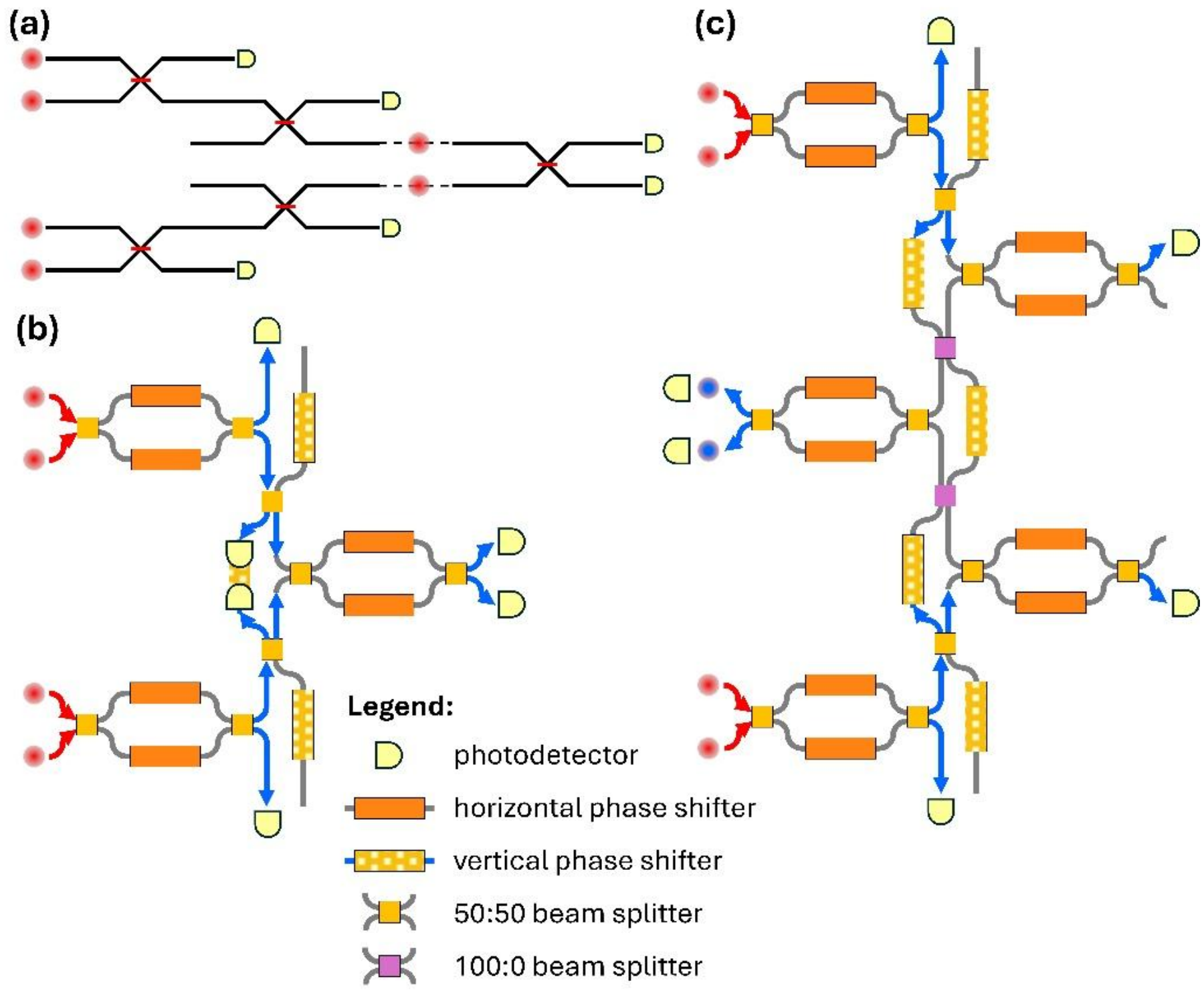

**Figure 4**. (a) A schematic of the linear optical circuit which implements two copies of the $n = 2$ purification circuit. (b, c) An equivalent of the above circuit implemented using a recirculating “bricks” mesh architecture.

Another example is illustrated in **Fig. 4a** which shows a schematic of the linear optical circuit consisting of cascaded interferometers organized in the tree configuration which implements two copies of the $n = 2$ purification circuits that perform quantum interference between the two purified outputs. Here, four streams of photons couple to different spatial modes. Each copy of the purification circuit consists of two beam splitters where each circuit provides the purified photon at the output. Then, the purified photons from each copy interfere in a final beam splitter. Finaly, photons are directed to six photodetectors to measure the output configurations. The experimental data show [**31**] that such a purification circuit provides a significant improvement in the photon indistinguishability, thereby enabling the realization of a successful purification protocol.

The circuit depicted in **Fig. 4a** is based on cascaded HOM interferometers realized in a feed-forward network using the tree configuration. Consequently, the system necessitates the implementation of four beam splitters, which are arranged in three layers. In comparison, the implementation of the aforementioned circuit in a recirculating "bricks" mesh architecture necessitates only two (**Fig. 4b**) or even one layer (**Fig. 4c**), thereby reducing the circuit's footprint and concurrently diminishing the propagation losses of photons. The capacity to launch and detect signals in all four directions, in conjunction with the ability of signals to propagate in any direction, facilitates the implementation of systems that would not be realizable using feed-forward networks.

**Multiphoton Fourier interference**

Most of the photon distillation protocols suffer from high resource cost of computation as each round of the protocol only slightly reduces the indistinguishability error. However, as has been previously demonstrated in numerous studies [**4, 5, 25, 26**], the quantum interference effects present in multimode interferometers may lead to the suppression of a significant proportion of the output configurations, contingent upon the implementation of a particular unitary transformation and the symmetry of the input state. And this, in turn, can lead to a significant reduction in the resource cost of computation.

One of the solutions for measuring genuine $n$-photon indistinguishability (GI) [**4**] is based on a Quantum Fourier transform (QFT) interferometer where for fully indistinguishable inputs certain output configurations are suppressed as a result of zero-transmission laws [**5, 26, 29, 34**]. It was originally proposed by Tichy [**35**] to reduce the resource costs and utilize the multiphoton interference in Fourier matrices. Tichy's seminal work on the Hong-Ou-Mandel effect for $n$ photons established foundational principles concerning the suppression laws that emerge when a fully indistinguishable input state is passed through a multi-port balanced beamsplitter. Crucially, the quantum operator employed in this context takes the form of the Quantum Fourier Transform (QFT), which is a crucial element in the development of quantum information theory. The corresponding $m \times m$ unitary matrix associated with this transformation is defined as:

$$QFT_m = \frac{1}{\sqrt{m}} \begin{bmatrix} 1 & 1 & \dots & 1 \\ 1 & u^1 & \dots & u^{m-1} \\ 1 & u^2 & \cdots & u^{2(m-1)} \\ \vdots & \vdots & \ddots & \vdots \\ 1 & u^{m-1} & \cdots & u^{(m-1)(m-1)} \end{bmatrix} \tag{1}$$

where $u = e^{2\pi i/m}$. These general suppression rules are known as the zero-transmission laws (ZTL). For a general case of $n = m$ (one photon per one mode) fully indistinguishable input state

$$Q(\vec{s}) \neq 0 \Longrightarrow \langle s|Q\hat{F}T|r\rangle = 0 \tag{2}$$

where $Q\hat{F}T$ represents the action of $QFT_m$ interferometer in Fock space and $Q$ is the $Q - value$ of an output configuration $\vec{s}$ for an input state $\vec{r}$ of $n$ indistinguishable photons [**25**]. In consequence, the suppression of many output configurations is observed when feeding this device with multi-photon states of a specific symmetry due to discrete many-particle interference.

For a Fourier transformation, the $m$-mode matrix elements of the optical transformations $U$ are expressed as:

$$U_m = \frac{1}{\sqrt{m}} e^{i2\pi(j-1)(k-1)/m} \tag{3}$$

where $m$ is the number of modes and optical transformations $U$ are implemented as a linear transformation acting on the bosonic modal creation operators

$$\left(\hat{s}_1^\dagger, \dots, \hat{s}_m^\dagger\right)^T \longmapsto U\left(\hat{r}_1^\dagger, \dots, \hat{r}_m^\dagger\right)^T \tag{4}$$

with $U$ being a unitary matrix. Such transformations are characterized by a high degree of symmetry what offers various attractive properties that can find application in quantum signal processing. As $U$ conserves photon number $n$, for the partial measurements in photon number basis for the conditions $\sum_{i=1}^{m-1} n_i = m - 1$, the state in the remaining mode is projected onto a single photon with an internal state $\rho(\epsilon')$ [**Fig. 3a**]. Where a single, fully indistinguishable photon is present in each input mode, the output distribution of a Fourier matrix obeys the zero-transmission law which results in the suppression of all output configurations not fulfilling the equation

$$\sum_{j=1}^{n} s_j \bmod(m) = 0 \tag{5}$$

where $s_j$, $s = (s_1, \dots, s_m)$, is the output mode of the $j^{th}$ photon corresponding to a particular measurement $|n_1, \dots, n_m\rangle\langle n_1, \dots, n_m|$. Such outcomes are referred to as allowed outcomes that can be efficiently predicted. In comparison, distinguishable photons do not obey the zero-transmission law (ZTL), what makes in this way the Fourier matrices an attractive probe for photon distinguishability.

The operation principle for a Fourier transformation is very similar to the one presented above for the cascaded interferometer (**Fig. 3a**) where the product state $\rho(\epsilon)^{\otimes n}$ is transformed by a $m$-mode Fourier transform $U$ into a multiphoton entangled state (**Fig. 5 and 6**). A given partial measurement of the multiphoton output state in the first $m-1$ auxiliary modes is indicative of the presence of a single photon state $\rho(\epsilon')$ in mode $m$ with reduced error $\epsilon' < \epsilon$. In **Fig. 5a** we have $\epsilon' = \epsilon/4$ for a single photon state with 3 auxiliary modes. Subsequently, the product state $\rho(\epsilon')\otimes\rho(\epsilon)$ is subjected to interference by a balanced beam splitter (not shown in **Fig. 5a and 6a**). The reduced partial distinguishability error $\epsilon'$ is then extracted by first computing the new visibility $V' = Tr[\rho(\epsilon)\rho(\epsilon')] = (1-\epsilon)(1-\epsilon')$ for known $\epsilon$. In short, this means that $\epsilon'$ is extracted by evaluating the Hong-Ou-Mandel (HOM) visibility of a distilled photon and a non-distilled photon by interference in a 50:50 beam splitter. The counting statistics of the balanced beam splitter for a given herald measurement are computed using the formalism for the interference of partially distinguishable photons [**35**]. Consequently, a single unitary transformation matrix can be used to describe all possible transformations to simulate the interference of $n+1$ partially distinguishable photons. Here, it was assumed that $n = m$, i.e., one photon per one mode.

In the context of a Fourier transform executed as a pairwise mode interactions, the number of required interactions per mode in the circuit, and thus the necessary number of beam splitters and phase shifters, exhibits a scaling behavior that is proportional to the logarithm of the number of modes involved, $O(m/2\log_2 m)$. This represents a significant reduction compared to the most general decomposition, in which the number of components scales with $O(m^2)$. Barak and Ben-Aryeh [**36**] developed such an efficient scheme called the quantum fast Fourier transform (qFFT) that allows for a significant reduction in number of optical elements on a photonic platform. This scheme is the quantum analogue of the fast Fourier transform (FFT), a well-known classical algorithm for efficiently calculating the discrete Fourier transform (DFT) that is known as the Cooley-Tukey algorithm [**37**].

The scheme proposed by Barak and Ben-Aryeh [**36**] provides an efficient procedure to implement linear optical transformations with a reduced number of elements. The resulting reduction is a direct consequence of the Cooley-Tukey algorithm which decompose a $2^n \times 2^n$ unitary matrix into $n$ steps of $2^{n-1}$ discrete transformations, thus, providing $n2^{n-1}$ qubit transformations.

In comparison, any unitary matrix of dimension $d$, here corresponding to the number of modes and thus $d = m$, can be decomposed to a multiplication of $d(d-1)/2$ unitary matrices that can be implemented using at most one beam splitter and one phase shifter acting on two qubit components. Thus, any $n$ qubit unitary transformation can be implemented by combination of $2^{n-1}(2^n - 1)$ beam splitters and phase shifters.

**Fig. 5b** shows a two-qubit optical DFT ($n = 2$) circuit realized with a recirculating "bricks" mesh architecture (**Fig. 5c**) in which the QFT is decomposed into a sequence of $n = 2$ partial transformations where each consist of $2^{n-1} = 2$ interactions between pairs of modes. Thus, a transformation matrix for the input operators is decomposed on the multiplication of two matrices in which the first matrix represents the first stage of the Cooley-Tukey algorithm (blue area in **Fig. 5a**) and the second matrix represents the second stage (red area in **Fig. 5a**). In consequence, the DFT circuit is composed of two 2-modes DFT circuits consisting of a beam splitter and phase shifter where the output of each is connected to the output of the other through another beam splitters and phase shifters. In consequence, only 4 pairs of beam splitters and phase shifters are required to perform transformations compared to 6 for a feed-forward mesh arranged in both Reck and Clements schemes.

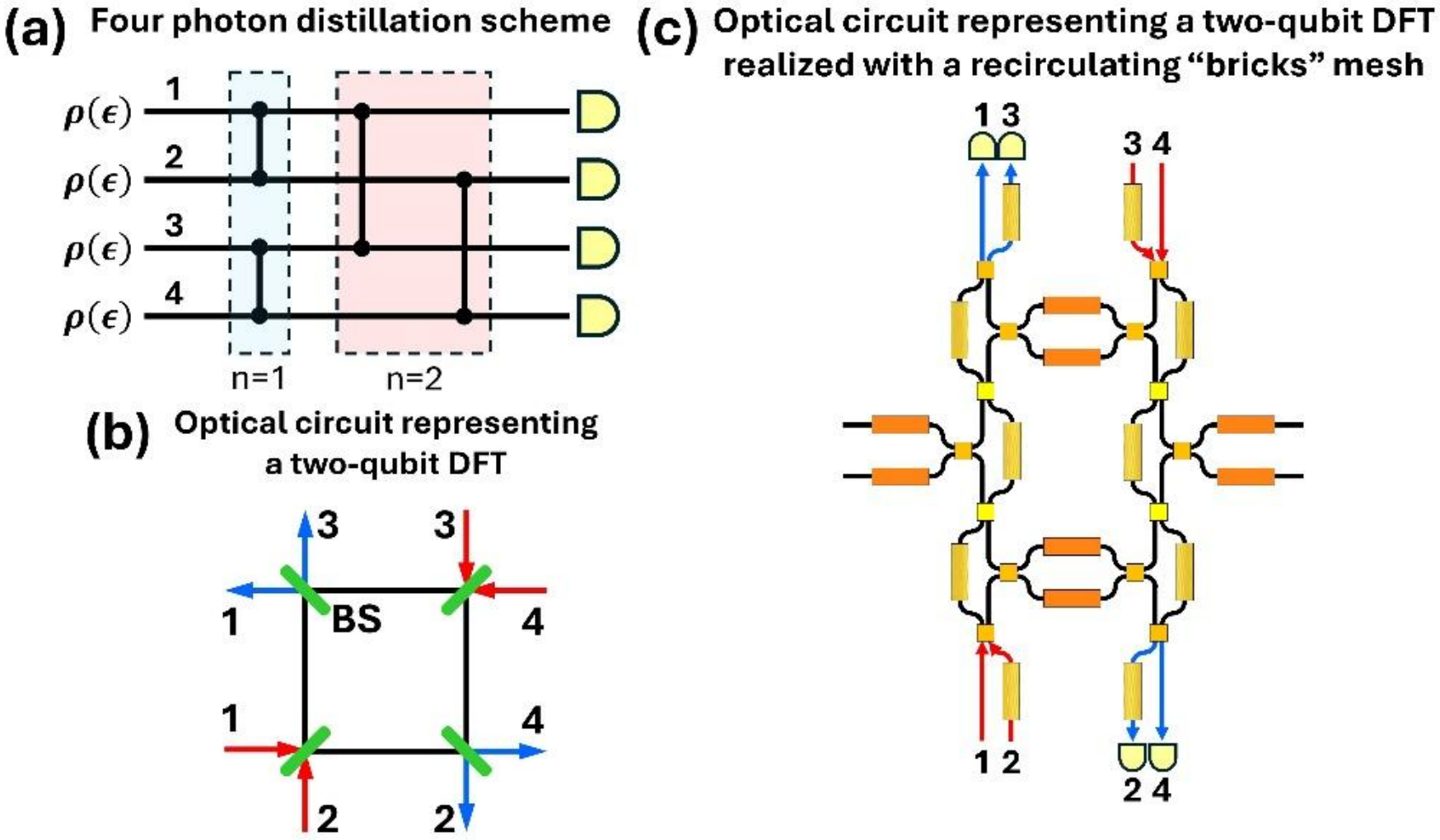

**Figure 5**. (a) Four mode DFT with a sequence of $n = 2$ partial transformations and (b) optical circuit representing a two-qubit DFT. (c) The corresponding two-qubit DFT realized with a recirculating "bricks" mesh.

In comparison, the 8-modes DFT circuit is presented in **Fig. 6a** with a corresponding connection schema for a three-qubit DFT presented in **Fig. 6b**. The equivalent circuit can be easily implemented in a recirculating "bricks" mesh architecture as shown in **Fig. 6c**. Using the Cooley-Tukey algorithm this transformation can be decomposed on the multiplication of three matrices QFT with each matrix representing a different stage in the Cooley-Tukey algorithm (**Fig. 6a**). Thus, the circuit is composed of two 4-modes DFT circuits where four outputs are combined with each other through a pair of beam splitter and phase shifter. Thus, a three-qubit DFT requires a minimum of 12 pairs of beam splitters and phase shifters compared to 28 for a feed-forward mesh arranged in both Reck and Clements schemes.

Each symmetric MZI implemented in a circuit can be composed of two 100:0 beam splitters to avoid additional interference, while two asymmetric MZIs can be designed using two 0:100 beam splitters to maintain the signal in the circuit. However, depending on the requirements, any other configuration of beam splitters is also possible.

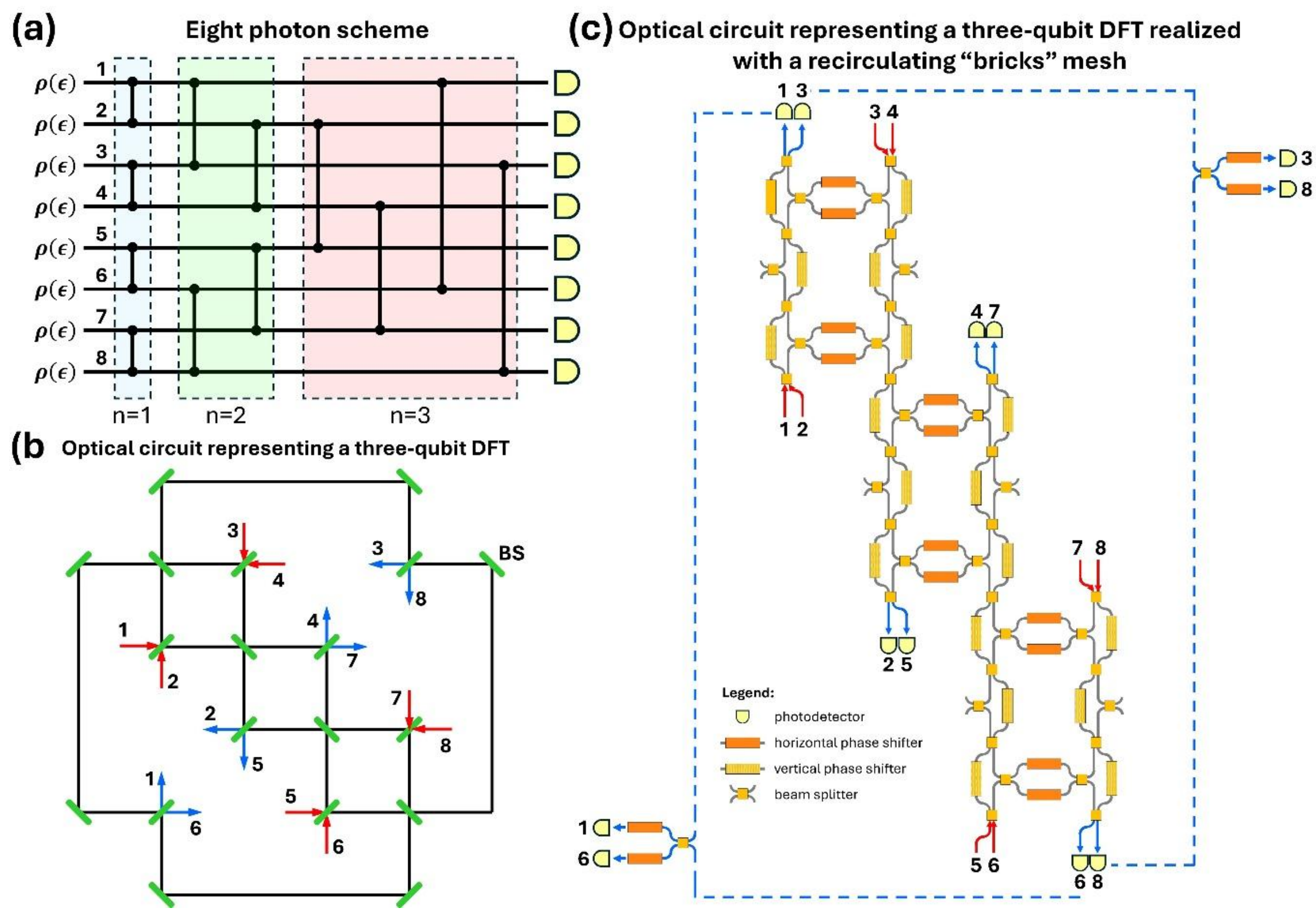

**Figure 6**. (a) Eight mode DFT with a sequence of $n = 3$ partial transformations and (b) optical circuit representing a three-qubit DFT. (c) The corresponding three-qubit DFT realized with a recirculating "bricks" mesh.

The prior arrangements are principally predicated on either Reck's or Clements mesh schemes and thus necessitate a non-planar disposition of the beam splitters and phase shifters into three-dimensional structures, which considerably increases the system's complexity. Both of those schemes are based on a feed-forward mesh, which significantly increases the system's optical depth and, consequently, losses. Furthermore, the decompositions of these schemes ignore possible symmetries of a given unitary and follows standard unitary matrix decomposition, so they are generally not optimized (the expected running time is $O(m^2)$, and scales with the number of independent parameters).

In comparison, the proposed recirculating "brick" mesh architecture enables the two-dimensional integration of all necessary components due to its unique ability to propagate signals in all directions. The ability to propagate signals in any direction combined with the ability to configure any port as an input or output port allows one to consider all possible symmetries in the system directly leading to a reduction in the number of necessary transformations. This simultaneously reduces the number of optical elements to $O(m/2\log_2 m)$, thereby limiting propagation losses and increasing photon indistinguishability.

## Summary

The recirculating meshes, particularly "bricks" meshes, offer substantial advantages in quantum technologies, primarily concerning scalability, high reconfigurability, and reduced optical losses when compared to conventional, static, large-scale photonic circuits. Rather than constructing a voluminous, single-directional array of interferometers, a recirculating mesh enables photons to traverse a diminutive, programmable, and adjustable component on multiple occasions, thereby simulating a substantial, intricate, and complex unitary transformation.

This facilitates the implementation of distillation protocols that are characterized by significantly reduced computational resource costs, as well as the implementation of systems that would be unattainable with feed-forward networks. Furthermore, the programmable nature of these systems allows for the implementation of a vast and diverse set of unitary transformations, thereby enabling the exploration of different quantum states within a single hardware configuration. This capability is of paramount importance for the testing and evaluation of various quantum technologies.

**Acknowledgement**

The author is very thankful to Prof. D. G. Misiek for his support and very valuable suggestions.